\documentclass[aps,prx, 11pt, notitlepage, floatfix,superscriptaddress, nofootinbib]{revtex4-2}
\usepackage{amssymb}
\usepackage{amsmath}
\usepackage{amsfonts}
\usepackage{bbm}
\usepackage{graphicx}
\usepackage{braket}
\usepackage[colorlinks,linkcolor=blue,anchorcolor=blue,citecolor=blue,urlcolor=blue]{hyperref}
\usepackage[dvipsnames]{xcolor}
\usepackage{qtex}
\usepackage[titletoc]{appendix}

\graphicspath{{Figures/}}

\def\S{\mathcal{S}}
\def\tr{\text{tr}}
\newcommand{\Sm}[1]{\mathcal{S}^{(#1)}}
\newcommand{\Smd}[1]{\mathcal{S}^{(#1),\dagger}}

\begin{document}
\title{Dynamics of operator size distribution in q-local quantum Brownian SYK and spin models}
\begin{abstract} 
We study operator dynamics in Brownian quantum many-body models with $q$-local interactions. The operator dynamics are characterized by the time-dependent size distribution, for which we derive an exact master equation in both the Brownian Majorana Sachdev-Ye-Kitaev (SYK) model and the spin model for general $q$. This equation can be solved numerically for large systems. Additionally, we obtain the analytical size distribution in the large $N$ limit for arbitrary initial conditions and $q$. The distributions for both models take the same form, related to the $\chi$-squared distribution by a change of variable, and strongly depend on the initial condition. For small initial sizes, the operator dynamics are characterized by a broad distribution that narrows as the initial size increases. When the initial operator size is below $q-2$ for the Majorana model or $q-1$ for the spin model, the distribution diverges in the small size limit at all times. The mean size of all operators, which can be directly measured by the out-of-time ordered correlator,  grows exponentially during the early time. In the late time regime, the mean size for a single Majorana or Pauli operator for all $q$ decays exponentially as $t e^{-t}$, much slower than all other operators, which decay as $e^{-t}$. At finite $N$, the size distribution exhibits modulo-dependent branching within a symmetry sector for the $q \geq 8$ Majorana model and the $q \geq 4$ spin model. Our results reveal universal features of operator dynamics in $q$-local quantum many-body systems.
\end{abstract}
\author{Shenglong Xu}
\email{slxu@tamu.edu}
\affiliation{Department of Physics \& Astronomy, Texas A\&M University, College Station, Texas 77843, USA}
\maketitle

\tableofcontents

\section{Introduction}
Exciting experimental progress on quantum simulators~\cite{altman2021quantum} calls for a theoretical understanding of the universal long-time quantum many-body dynamics beyond the paradigm of quantum thermalization~\cite{deutsch1991quantum,srednicki1994chaos,rigol2008thermalization,polkovnikov2011colloquium}.  In a generic interacting many-body system, the quantum state undergoes unitary time evolution, and keep evolving even after the local density matrix reaches equilibrium. Beyond thermalization, the universal aspect of the unitary dynamics manifests in non-local degree freedom, through entanglement generation~\cite{kim2013ballistic, ho2017entanglement,nahum2017quantum} in the Schrodinger picture and operator growth in the Heisenberg picture~\cite{nahum2018operator, von2018operator, parker2019universal}.

In the Heisenberg picture, operator growth refers to how a simple local operator, such as a single Pauli or Majorana operator, becomes non-local under unitary dynamics. The size of the operator is directly related to the out-of-time ordered correlator (OTOC)~\cite{larkin1969quasiclassical}, which is commonly used as a diagnostic tool for scrambling dynamics~\cite{lewis2019dynamics,xu2022scrambling}. The OTOC can be seen as a type of Green's function defined on a 4-fold Keldysh contour and has been extensively studied in a variety of many-body models and measured on various quantum simulation platforms~\cite{mi2021information, landsman2019verified, garttner2017measuring,braumuller2021probing, wang2021verifying,sanchez2020perturbation}. A typical Heisenberg operator involves a superposition of operators with different sizes and is thus characterized by a distribution of size. The operator size distribution provides a more detailed probe of scrambling dynamics than the mean size, revealing the full spectrum of operator growth during unitary time evolution~\cite{zhou2019operator, qi2019quantum, roberts2018operator, lucas2020nonperturbative, zhang2023operator, yao2024notessolvable}, as well as in open systems~\cite{schuster2023operator, zhang2023dynamical}. Furthermore, the operator size distribution is directly tied to the fidelity in quantum many-body teleportation protocols~\cite{hayden2007black, yoshida2017efficient, gao2021traversable, brown2019quantum, nezami2021quantum, schuster2021many} and thus has a significant quantum information application.  Recently, it has been proposed that the operator size distribution can be directly measured in quantum quench experiments~\cite{qi2019measuring}.

We study the operator size dynamics using a class of quantum many-body models known as the Brownian model~\cite{lashkari2013towards, xu2019locality,knap2018entanglement,saad2018semiclassical, sunderhauf2019quantum, jian2021note, agarwal2021emergent, stanford2022subleading, guo2024complexity, swann2023spacetime, jian2022linear, vardhan2024entanglement}, which is driven by external dephasing noise and can be regarded as a continuous-time version of random quantum circuits~\cite{fisher2022random}. Specifically, we consider the Brownian Majorana Sachdev-Ye-Kitaev (SYK) model~\cite{saad2018semiclassical} and spin models with \( q \)-body interactions. It is known that, after a random average, one can derive an exact master equation describing the operator size distribution~\cite{xu2019locality, zhou2019operator, jian2021note}, which can be simulated classically for large system sizes.

In this work, we introduce a new method to simplify the derivation of the master equation and extend previous results to arbitrary $q$-body interactions. After the random average, the real-time dynamics of the Brownian model is governed by the imaginary time evolution of an emergent Hamiltonian on multiple replicas. We show that the emergent Hamiltonian can be mapped to a classical master equation by a similar transformation. Based on the master equation, we also develop an analytical theory of operator size dynamics in the large \( N \) limit for arbitrary \( q \) and initial operators, by combining the exact early-time results of the master equation with the long-time continuum limit. Remarkably, we find that the size distributions for the Majorana model and the spin model take the same form, connecting to the \(\chi\)-squared distribution $f$ by a change of variable.
\begin{equation}
    p(x,t) = f_{m_0/\lambda}(z)\frac{dz}{dx}, \quad f_{m_0/\lambda}(z) = \frac{1}{\Gamma(m_0/\lambda)}e^{-z}z^{m_0/\lambda-1}
\end{equation}
where $m_0$ is the size of the initial operator and $\lambda$ equals $q-2$ and $q-1$ for the Majorana and the spin model, respectively. The new time-dependent variable $z$ is related to the original size density $x$ as
\begin{equation}
    z = x^{eq}\frac{e^{\lambda(t^* - t)}}{\lambda}(-1+(1-x/x^{eq})^{-\lambda}), \quad t^* = \frac{1}{\lambda}\left( \log N - \log\lambda\right)
\end{equation}
where $t^*$ is the scrambling time and $x^{eq}$ is the equilibrium value of the mean size. It equals $1/2$ in the Majorana model and $3/4$ in the spin model. Once $\lambda$ and $x^{eq}$ are fixed, the operator dynamics of the Majorana model and the spin model are characterized by the same size distribution at all times in the large $N$ limit.

Through both numerical simulations and analytical methods, we demonstrate that the size of the initial operator significantly affects the long-time dynamics:
\begin{itemize}
    \item Due to the \(\chi\)-squared distribution, the Heisenberg operator of a small initial size \(m_0\) is characterized by a broad distribution.
    \item When \(m_0 < \lambda\), the size distribution retains a sharp peak at small sizes that diverges as $x^{m_0/\lambda-1}$ at all times. 
    \item The width of the distribution decreases as \(m_0^{-1/2}\) as the initial size increases.
    \item The mean size of all operators grows exponentially as $e^{\lambda t}$ in the early time, and the rate, known as the Lyapunov exponent, equals $\lambda$ defined above. 
    \item The mean size of a single Majorana operator or a single Pauli operator decays as \(t e^{-t}\) in the late time for arbitrary $q$, independent of \(\lambda\), while all other operators decay exponentially as \(e^{-t}\).
\end{itemize}  
Our results reveal new universal features of operator dynamics in \(q\)-local quantum many-body systems.

\section{Operator size distribution}
\label{sec:osd}
We first review the standard definition of operator size for spin systems and Majorana systems. 
\subsection{Spin systems}
In a system of $N$ spin 1/2, any operator can be expanded in a complete basis of operators called Pauli strings, 
\begin{equation}
    O = \sum \alpha(\S)\S.
\end{equation}
A Pauli string $\mathcal{S}$ is a tensor product of $N$ local Pauli operators $\sigma^x$, $\sigma^y$ and $\sigma^z$, and identity operator $I$,
\begin{equation}
    \S = s_1 s_2 \cdots s_N, \quad s_i \in \{\sigma^x_i, \sigma^y_i, \sigma^z_i, I_i\}
\end{equation}
There are a total of $4^N$ Pauli strings. These Pauli strings satisfy the following orthonormal relation,
\begin{equation}
   \frac{1}{2^N} \text{tr}(\S \S'^\dagger) = \delta_{\S \S'} 
\end{equation}
Therefore, the normalization of $W$ is related to the expansion coefficient as,
\begin{equation}
    \frac{1}{2^N} \tr(O O^\dagger) = \sum\limits_{\S} |\alpha(\S)|^2.
\end{equation}
This norm does not change during unitary Heisenberg time evolution and is set to 1 conventionally. As such, $\sum |\alpha(\S,t)|^2$ can be interpreted as a time-dependent probability distribution.

The operator size can be understood as the support of an operator, i.e., how many sites this operator acts on.  Therefore, a Pauli string has a definite operator size, which equals the number of local Pauli operators in the string. An operator $O$ is a superposition of Pauli strings with different sizes. We can define an operator size distribution from the expansion coefficient by grouping all Pauli strings of the same size. Denote $\S^{(m)}$ as a Pauli string of size $m$. Then the operator size distribution is,
\begin{equation}
    P(m) = \sum \limits_{\S^{(m)}} |\alpha(\S^{(m))}|^2,
\end{equation}
and we have $\sum P_m = 1$ from the normalization of the Heisenberg operator.
An initial local operator has a small size. As time increases, the operator spreads over the system, and the operator size distribution shifts toward large sizes, signaling the operator growth. In the long run, assuming every Pauli string becomes equally probable, the operator size distribution becomes a binomial distribution
\begin{equation}
    \lim \limits_{t\rightarrow \infty} P(m,t) = \frac{1}{4^N}3^m\binom{N}{m} 
\end{equation}
Of course, this is without considering any conserved quantity that brings constraints on the operator dynamics. 

Formally, the operator size is measured by a super operator $\mathbb{M}=\sum \mathbb{M}_i$, where each term acts on a regular operator as
\begin{equation}
    \mathbb{M}_i (O) = \frac{3}{4}O -\frac{1}{4} (\sigma_i^x O \sigma_i^x + \sigma_i^y O \sigma_i^y + \sigma_i^z O \sigma_i^z). 
\end{equation}
One can verify that $\mathbb{M}_i (I_i) = 0$ and $\mathbb{M}_i(\sigma^s_i) = \sigma^s_i$. Hence $\mathbb{M}_i$ checks whether the local operator at site $i$ is an identity or a Pauli operator, and $\mathbb{M}$ counts the number of local Pauli operators, i.e., the size of the operators. Any Pauli string $\S$ is an eigenoperator of the super operator $\mathbb{M}$ and the eigenvalue is its size,
\begin{equation}
    \mathbb{M}(\S^{(m)}) = m \S^{(m)}.
\end{equation}
The expectation value of size for a general Heisenberg operator $O(t)$ is
\begin{equation}
    \langle m \rangle = \frac{1}{2^N}\tr(O(t)^\dagger \mathbb{M}(O(t))) =\sum\limits_m P(m,t)m
\end{equation}
Plugging in $\mathbb{M}$ gives,
\begin{equation}
    \langle m \rangle = \frac{1}{4}\sum_{i, s}\left(1-\frac{1}{2^N}\tr(O^\dagger(t)\sigma^s_i O(t) \sigma^s_i)\right)
\end{equation}
Each term in the summation is an out-of-time-ordered correlation function~(OTOC)~\cite{larkin1969quasiclassical}. 

\subsection{Majorana systems}
In fermionic systems without symmetry, such as charge conservation, it is more convenient to use Majorana operators. In a system of $N$ Majoranas, a complete basis of operators consists of Majorana strings $\S$, which are products of local Majorana operators
\begin{equation}
    \S =  \chi_{i_1} \cdots \chi_{i_m}, \ i_1<\cdots<i_m
\end{equation}
where $\chi_i$ is the Majorana operator on site $i$. These $2^N$ Majorana operators satisfy the anticommutation relation,
\begin{equation}
    \{\chi_i, \chi_j \} = 2 \delta _{ij}.
\end{equation}
The coefficient $2$ is introduced so that $\chi_i^2=1$ and each Majorana string has the same norm. The Majorana strings obey the following orthonormal relation
\begin{equation}
    \frac{1}{2^{N/2}}\tr(\S^\dagger \S') = \delta_{\S\S'}
\end{equation}

The operator size in Majorana systems, closely following that in spin systems, simply counts the number of Majorana operators in the string. The operator size distribution is the probability of all Majorana strings of size $m$ appearing in the operator expansion. In this case, the size super operator is
\begin{equation}
    \mathbb{M} = \sum_i\mathbb{M}_i, \quad \mathbb{M}_i(O) = \frac{1}{2} O - \frac{1}{2} \chi_i \mathcal{P} O \mathcal{P} \chi_i
\end{equation}
where $\mathbb{M}_i$ checks whether the operator locally contains $\chi_i$. The parity operator $\mathcal{P}$ is included to cancel out the extra sign due to the anti-commutation relation between the Majorana. 

Similar to the spin system, the expectation of the size for a general Heisenberg operator in the Majorana system is directly related to the OTOC,
\begin{equation}
    \langle m \rangle = \frac{1}{2^N}\tr(O(t)^\dagger \mathbb{M}(O(t))) =\frac{1}{2}\sum_{i}\left(1-\frac{1}{2^{N/2}}\tr(O^\dagger(t)\chi_i\mathcal{P} O(t)\mathcal{P} \chi_i)\right)
\end{equation}

\section{Conditions for classical simulation of operator size distribution }
\label{sec:condition}
The goal of this work is to study universal features of the operator size distribution in interacting $q$-local quantum many-body systems. 
In general, calculating the size distribution for systems requires solving Schrodinger's equation in an exponentially large Hilbert space, which is a formidable task, even though the size distribution only has $N+1$ components. Therefore one usually can only study operator size distribution in small systems using exact diagonalization or in some large $N$ semi-classical limit. However, there exists a class of random quantum many-body model, in which by taking advantage of the random average, one can describe the dynamics of operator size distribution by an exact stochastic process which can be simulated classically~\cite{chen2019quantum,xu2019locality}. We also note that operator dynamics can be solved exactly in certain strongly interacting non-random quantum systems beyond exact diagonalization, such as dual unitary circuits~\cite{bertini2019exact, piroli2020exact, claeys2020maximum}. 

In the following, we discuss sufficient conditions for the classical description to emerge in random quantum systems. 
In both Majorana and spin systems, the operator size distribution for a Heisenberg operator $O(t)$ is
\begin{equation}
    P(m,t) =\frac{1}{D^2}\sum_{\S^{(m)}} |\tr(\S^{(m)}O(t))|^2,
\end{equation}
where $\S^{(m)}$ are Majorana/Pauli strings of size $m$ defined in Sec.~\ref{sec:osd}, and $D$ is the Hilbert space dimension.
It is common in the literature to simplify the notation using replica notation by operator-state mapping. Define the super unitary operator on four replicas $\mathbb{U}=U\otimes U^* \otimes U\otimes U^*$. Then $P(m)$ is the overlap between two operator-states on four replicas,
\begin{equation}
    P(m,t) = \frac{1}{D^2}\sum_{\Sm{m}}\bra{\Sm{m}\otimes \Smd{m}}  \mathbb{U}(t) \ket{O \otimes O^\dagger}.
\end{equation}
We also introduce the following tensor network notation for the size distribution,
\begin{equation}
P(m,t) = \frac{1}{D^2} \ \
    \begin{qtex}[2.2]
    \grid{2}{0;4,0}
    \gate[][center:$U$]{0;2;2,1}
    \gate[][center:$U^*$]{1;2;2,1}
    \connect{0,0}{1,0}
    \connect{2,0}{3,0}
    \connect{0,2}{1,2}
    \connect{2,2}{3,2}
    \connect[dashed][0.2]{0.5,2}{2.5,2}
    \qubit[][center:$O$]{0.5,0}
    \qubit[][center:$O^\dagger$]{2.5,0}
    \qubit[][center:$\S^{(m),\dagger}$]{0.5,2}
    \qubit[][center:$\S^{(m)}$]{2.5,2}
    \end{qtex}
\end{equation}
The dashed line presents summing over $\S^{(m)}$ for a fixed $m$.
Using the completeness relation of the operator basis $\S$, we have
\begin{equation}
    P(m,t) = \frac{1}{D^4}\sum\limits_{\Sm{m},\S',\S''} \bra{\Sm{m}\otimes \Smd{m}}\mathbb{U}(t)\ket{\S'\otimes \S''^\dagger}\bra{\S' \otimes \S''^\dagger}O \otimes O^\dagger \rangle.
\end{equation}
We state two conditions on the 4-replica super operator $\mathbb{U}$ to derive a classical stochastic equation for the operator size distribution:
\begin{enumerate}
    \item The super unitary operator $\overline{\mathbb{U}}$ after disorder average is closed in the  Hilbert space spanned by $\ket{\S \otimes \S^\dagger}$,
    \begin{equation}
        \bra{\S\otimes \S^\dagger} \overline{\mathbb{U}} \ket{\S' \otimes \S''^\dagger} =   \bra{\S\otimes \S^\dagger} \overline{\mathbb{U}} \ket{\S' \otimes \S'^\dagger} \delta_{\S'\S''}.
    \end{equation}
    Without random average, $\mathbb{U}$ acts on the four replicas independently, and this condition is never satisfied. Below, we use $\mathbb{U}$ to represent the super unitary operator after random average.
    
    \item The matrix element of the super unitary operator $\bra{\S \otimes \S^\dagger} \mathbb{U} \ket{\S' \otimes \S'^\dagger}$ does not depend on the specific operator strings but only on their sizes.
    As a result, $\mathbb{U}$ is closed in a much smaller Hilbert space spanned by the basis called operator size basis,
    \begin{equation}
    \label{eq:operatorsizebasis}
        \ket{\psi_m} = \frac{1}{D D(m)^{1/2}}\sum \limits_{\Sm{m}}\ket{\Sm{m}\otimes \Smd{m}},
    \end{equation}
    where $D(m)$ is the number of basis operator strings of size $m$.
\end{enumerate}
By applying the first condition,  $P(m,t)$ can be significantly simplified as follows,
\begin{equation}
\begin{aligned}
    P(m,t) & = \frac{1}{D^4}\sum\limits_{\Sm{m},\S'} \bra{\Sm{m}\otimes \Smd{m}}\mathbb{U}(t)\ket{\S' \otimes \S'^\dagger}\bra{\S' \otimes \S'^\dagger}O \otimes O^\dagger \rangle \\
    & = \frac{1}{D^4}\sum\limits_{\Sm{m},\Sm{m'},m'} \bra{\Sm{m}\otimes \Smd{m}}\mathbb{U}(t)\ket{\Sm{m'}\otimes \Smd{m'}}\bra{\Sm{m'}\otimes \Smd{m'}}O \otimes O^\dagger \rangle
\end{aligned}
\end{equation}
Then applying the second condition gives rise to a stochastic equation
\begin{equation}
    P(m,t) = \sum_n A_{m,m'}(t) P(m',t=0), \quad  A_{m,m'}(t) =\sqrt{\frac{D(m)}{D(m')}} \bra{\psi_m} \mathbb{U}(t) \ket{\psi_m'} 
\end{equation}
The matrix $A$ is related to $\mathbb{U}$ by a similar transformation. One can verify that $\sum_m A_{m,n}(t)=1$ and $A_{m,n}\geq 0$ and it is indeed a stochastic matrix. 

We note that the first condition above on its own is sufficient to derive a stochastic equation for the operator dynamics. The dimension of the resulting stochastic matrix is exponentially large as it applies to all the components $|\alpha(\S)|^2$. Then the stochastic equation cannot be solved straightforwardly but nevertheless can be sampled using classical Monte Carlo~\cite{chen2019quantum}. The second condition is to reduce the dimension of the stochastic matrix from exponential to linear in system size.

\section{ Brownian models and exact master equations for operator size distribution}
\subsection{General structure}
There are many ways to generate a random unitary ensemble, such as through random quantum circuits where each gate is selected from a random ensemble, or using a time-dependent Hamiltonian driven by noise. 
Let us focus on the latter case, and consider a class of models known as the quantum Brownian model~\cite{lashkari2013towards}, described by,
\begin{equation}
\label{eq:generalBrownian}
    H(t) = \sum_A J_A(t) X_A, \overline{J_A(t)}=0, \overline{J_A(t)J_{A'}(t')} = J \delta_{A,A'}\delta(t-t')
\end{equation}
for some Hermitian operators $X_A$ and the corresponding random coefficients $J_A(t)$, which are uncorrelated for different operators and times. The unitary time evolution generated by this Hamiltonian is $U=\mathcal{T}\exp(-i \int H(t) dt)$. Since the Hamiltonian is uncorrelated, one can take the random average independently at each time slice. As a result, the super unitary operator becomes an imaginary time evolution operator for an effective static Hamiltonian acting on four replicas~\cite{sunderhauf2019quantum, agarwal2021emergent, guo2024complexity},
\begin{equation}
\label{effH}
    \mathbb{U} = \exp(-\mathbb{H}t), \quad \mathbb{H} = \frac{J}{2}\sum_A (X_A^{a} -X_A^{*,b} - X_A^{c} -X_A^{*,d})^2,
\end{equation}
where the superscript $a\sim d$ labels each of the four replicas.
One can derive this result by expanding the infinitesimal time evolution $\exp(-i H dt)$ to the second order and then random averaging the coefficients over the four replicas. The effective Hamiltonian has a few nice properties worth mentioning here. First, it is positive because the super unitary operator $\mathbb{U}$ must be bounded at late times. Second, the effective Hamiltonian has at least two zero-energy ground states that each term shares and thus is frustration free. These two ground states are inherited from the identity operator, which remains static under unitary time evolution. Third, the Hamiltonian is invariant under permutations within forward/backward replicas or exchanging forward and backward replicas combined with time reversal~\cite{bao2021symmetry}.

Imposing the two conditions discussed in Sec.~\ref{sec:condition} on the super unitary operator is equivalent to imposing them on the effective Hamiltonian. If these conditions hold, the dynamics of the operator size distribution are described by a master equation whose dimension is linear in the system size,
\begin{equation}
\label{eq:generalMaster}
    \frac{d}{dt} \vec P(t) = A \vec P(t),\quad  \quad  A_{m,m'} =-\sqrt{\frac{D(m)}{D(m')}} \bra{\psi_m} \mathbb{H} \ket{\psi_{m'}}.
\end{equation}
The matrix $A$ and the effective Hamiltonian in the operator size basis are related by a simple similar transformation.

\subsection{Exact master equations for q-local Brownian Majorana model and spin model}
\label{sec:master}
\subsubsection{Models}
We consider all-to-all connected $q$ local Brownian Hamiltonian for both Majoranas~\cite{saad2018semiclassical} and spins. 
For Majoranas, the operators $X_A$ in the general Hamiltonian Eq.~\eqref{eq:generalBrownian} are chosen to be all Majorana strings of size $q$, and there are $\binom{N}{q}$ terms in the Hamiltonian
\begin{equation}
    H(t) = \sum  i^{q/2} J_{i_1\cdots i_q}(t) \chi_{i_i} \chi_{i_2}\cdots \chi_{i_q}
\end{equation}
Recall that $\chi_i$ is the Majorana operator acting site $i$. These operators satisfy the anicommutation relation $\{\chi_i, \chi_j \}=2\delta_{ij}$. The factor of 2 is introduced so that every Majorana string has the same norm.

For the spin model, the operators $X_A$ are all Pauli strings of size $q$, and there there are $3^q\binom{N}{q}$ terms in the Hamiltonian 
\begin{equation}
    H(t)  = \sum J^{s_1\cdots s_q}_{i_1\cdots i_q}(t) \sigma^{s_1}_{i_1} \sigma^{s_2}_{i_2}\cdots \sigma^{s_q}_{i_q} 
\end{equation}
where $\sigma^s$ for $s=1\sim 3$ are the usual Pauli matrices. The spin model with $q=2$ was introduced in~\cite{lashkari2013towards}.

Now we show that the effective Hamiltonian for the Brownian Majorana and spin model has a closed subspace spanned by the operator basis $\ket{\S\otimes \S}$, thus satisfying the first condition. Notice that the operators $X_A$ appearing in the Brownian Hamiltonian are also Majorana/Pauli strings of length $q$. Therefore $X_A$ commutes or anticommutes with $\S$, and $X_A^2=I$. The effective Hamiltonian acts on the state $\ket{\S\otimes \S^\dagger}$ as,
\begin{equation}
\label{eq:Xa}
    \frac{1}{2}(X_A^{a} - X_A^{*,b} + X_A^{c} -X_A^{*,d})^2 \ket{\S\otimes \S^\dagger} = 4 \ket{\S\otimes \S^\dagger} - 4 \ket{X_A\S\otimes \S^\dagger X_A}
\end{equation}
when $X_A$ anticommutes with $\S$ and zero when they commute. Observe that the product $X_A \S$ is also a Pauli/Majorana string. Therefore $\mathbb{H}$ satisfies the first condition. Furthermore, the effective Hamiltonian acts on all the operator strings of the same size equivalently, and therefore its matrix element only depends on the size of the operator strings, satisfying the second condition. Based on the discussion in Sec.~\ref{sec:condition}, these two conditions ensure that one can derive an exact master equation Eq.~\eqref{eq:generalMaster} describing the operator size dynamics in the Brownian majorana/spin model. We note that due to some emergent symmetry of the Majorana model~\cite{agarwal2021emergent}, one can obtain the entire spectrum of $\mathbb{H}$ beyond the operator size basis for very large system sizes. 

Now we derive the matrix elements $\bra{\psi_m}\mathbb{H}\ket{\psi_{m'}}$ to get the master equation. Since the matrix element $\bra{\Sm{m}\otimes \S^{(m)\dagger}} \mathbb{H} \ket{\Sm{m'}\otimes \S^{(m')\dagger}}$ only depends on $m$ and $m'$, we have
\begin{equation}
    A_{m,m'} =-\sqrt{\frac{D(m)}{D(m')}} \bra{\psi_m} \mathbb{H} \ket{\psi_{m'}} = -\frac{1}{D^2}\sum_{\Sm{m}}\bra{\Sm{m}\otimes \S^{(m)\dagger}} \mathbb{H} \ket{\Sm{m'}\otimes \S^{(m')\dagger}}.
\end{equation}
This is proportional to the number of operator strings of size $m$ generated by acting the effective Hamiltonian on a single string of size $m'$.

\subsubsection{Majorana master equation}
For simplicity, we first consider the Majorana model, where $X_A=(i)^{q/2}\chi_{i_1}\cdots\chi_{i_q}$. The operator $X_A$, which contains an even number of Majorana operators, anticommutes with $\Sm{m}$ when they share an odd number of Majorana operators, denoted $n$. There are a total of $\binom{m}{n}\binom{N-m}{q-n}$ such operators $X_A$. The size of $X_A \Sm{m}$ is $m+q-2n$. Then we obtain the exact master equation:
\begin{equation}
\label{eq:MajoranaEquation}
    \partial_t P(m) = \sum_{n\in odd,1\leq n < q}\gamma^n_{\chi} (m+2n-q) P(m+2n-q) -\gamma^n_{\chi}(m) P(m), \ \
\gamma^n_{\chi}(m) = 4 J_\chi \binom{N-m}{q-n}\binom{m}{n}.
\end{equation}
where $\gamma^n_{\chi}(m)$ is the transition rate from operators of size $m$ to operators of size $m+q-2n$. We set $J_\chi$, the variance of random coefficients in the Brownian Hamiltonian, to
\begin{equation}
\label{eq:Jchi}
J_\chi = \frac{N}{4q} \binom{N}{q}^{-1}.    
\end{equation}
This choice aligns with the standard SYK model~\cite{maldacena2016remarks}. From the perspective of the master equation, this normalization ensures that $\gamma^{n=1}(m)=m$, independent of $N$ and $q$ in the large $N$ limit. In particular $\gamma_\chi^{n=1}(1)=1$.

Due to fermionic parity conservation, the size change must be an even number. Furthermore, when $q/2$ is odd, the Hamiltonian is odd under time reversal $K$, which is simply complex conjugate, because of the phase factor $i^{q/2}$. As a result, the unitary time evolution commutes $K$, and therefore the Heisenberg operator must transform the same way under time reversal as the initial operator. The time reversal and parity conservation imply that the operator size change must be a multiple of 4 when $q/2$ is odd, as shown by the master equation. As a result of the symmetry consideration, the master equation is divided into 2 and 4 decoupled sectors for even $q/2$ and odd $q/2$, respectively.
We emphasize that the time reversal and parity conservation are not exclusive to the Brownian nature of the model but also apply to the static SYK model. 

We note that, while we focus here on the effective Hamiltonian $\mathbb{H}$ within the subspace spanned by the operator size basis $\ket{\psi_m}$, it is possible to study the entire spectrum of $\mathbb{H}$. This is because after disorder averaging the full Hilbert space splits into exponentially many subspaces, whose dimensions scale at most as $N^2$ due to an emergent SU(2)$\otimes$SU(2) algebra for general $q$~\cite{agarwal2021emergent}.

\subsubsection{Spin master equation}
Now we turn to the Brownian spin model.  To our knowledge, its master equation has only been derived for $q=2$ so far in the literature~\cite{lashkari2013towards, xu2019locality, zhou2019operator}. 
Without loss of generality,  consider a Pauli string $\Sm{m}$ that $m$ $\sigma^z$ on the first $m$ sites and identities in the remaining sites. First, $X_A$, a Pauli string of size $q$, needs to anticommute with $\Sm{m}$ so that Eq.~\eqref{eq:Xa} is not zero.
This requires that $X_A$ contains an odd number of $\sigma^x$ and $\sigma^y$ in the first $m$ site. In $X_A$, any $\sigma^z$ in the first $m$ site decreases the operator size while any Pauli operator beyond the first $m$ sites increases the operator size.
It is convenient to parameterize the size change as $q-2n+1$ for integer $n$ ranging from 1 to $q$. 
From combinatorics, the number of generated operators of size $m+q-2n+1$ is
\begin{equation}
\label{eq:spinRate}
  \gamma^n_{\sigma}(m) = 4J  \sum_{n_z} 2^{2n-2n_z-1}3^{n_z+q-2n+1}\binom{m}{n_z}\binom{N-m}{n_z+q-2n+1}\binom{m-n_z}{2n-2n_z-1}
\end{equation}
where $n_z$ is the number of $\sigma_z$ in the first $m$ site, $2n-2n_z-1$ is the number of $\sigma^x$ and $\sigma^y$ operator in the first $m$ site, and $n_z+q-2n+1$ is the number of Pauli operator beyond the first $m$ site. This series formally sums to a generalized hypergeometric function but can be evaluated efficiently for a given $q$. 
We choose the normalization of $J_\sigma$ as
\begin{equation}
\label{eq:Jsigma}
    J_\sigma = \frac{N}{8q 3^{q-1}}\binom{N}{q}^{-1}.
\end{equation}
With the chosen normalization of $J_\chi$ in Eq.~\eqref{eq:Jchi} and $J_\sigma$, the transition rates $\gamma^{(1)}(m)$ corresponding to the largest size increase are the same in both models. In the Majorana model, the largest size increase of $X_A\S^{m}$ is $q-2$, which occurs when $X_A$ shares a single Majorana with $\Sm{m}$. In the spin model, the largest size increase is $q-1$,  when $X_A$ and $\Sm{m}$ overlap on one site and have different Pauli operators on that site. 
The master equation of the spin model takes a similar form as the Majorana model but with the transition rate described above.
\begin{equation}
\label{eq:spinEquation}
    \partial_t P(m) = \sum_{1\leq n \leq q}\gamma^n_{\sigma} (m+2n-1-q) P(m+2n-1-q) -\gamma^n_\sigma(m) P(m).
\end{equation}
For odd $q$, the change of the operator size must be an even number because the Hamiltonian is odd under time reversal given by $T = K\prod \sigma^y_i$. As a result, the master equation is divided into two decoupled sectors for odd $q$.

\subsection{Large system numerics}

\begin{figure}
    \includegraphics{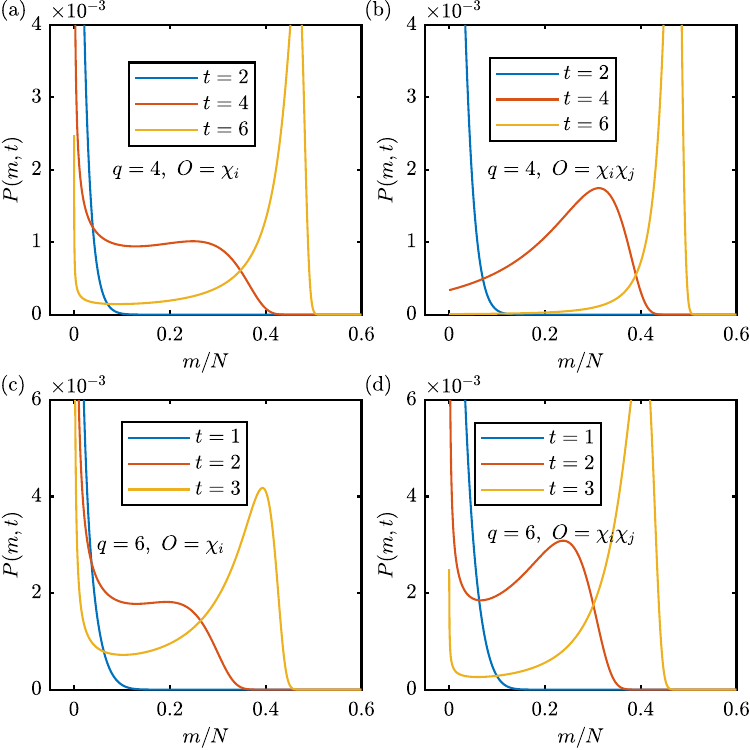}
    \caption{Numerical results of the Majorana operator size distribution for different initial operators and $q$ by directly simulating the exact master equation in Eq.~\eqref{eq:MajoranaEquation}. As time increases, the weight of the distribution shifts from the initial value close to 0 to the steady value at 1/2. (a) The initial operator is a single Majorana operator with size 1, and $q$ is 4. 
    The distribution displays two peaks, one at the origin and one at the steady value at $1/2$. (b) Same as (a) but for a double Majorana operator with size 2. (c) The initial operator is a single Majorana operator with size 1, and $q$ is 6. (d) Same as (c) but for a double Majorana initial operator. At $q=6$, both initial conditions lead to two peaks in the size distributions. }
    \label{fig:numericalMajorana}
\end{figure}

\begin{figure}
    \includegraphics{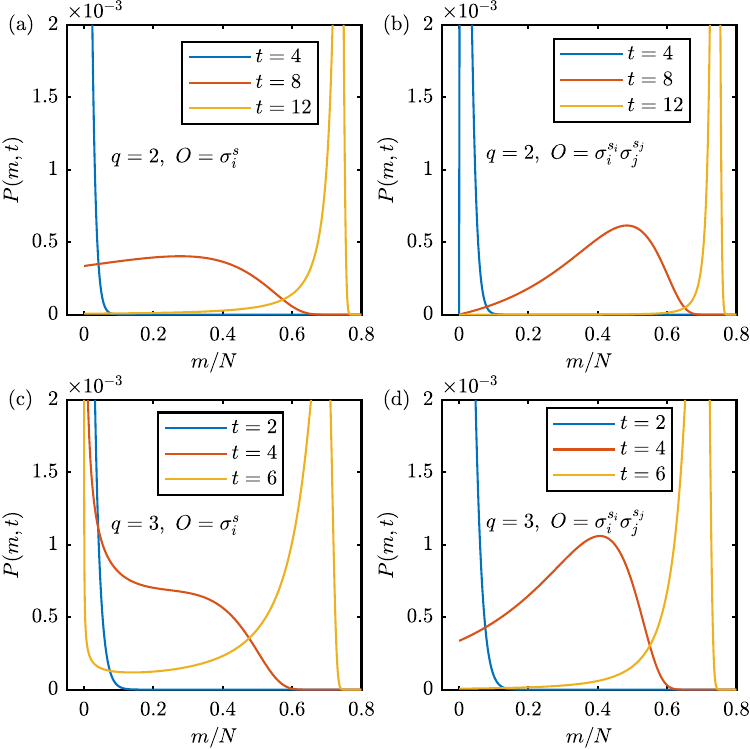}
    \caption{Numerical results of the spin operator size distribution for different initial operators and $q$ by directly simulating the exact master equation in Eq.~\eqref{eq:spinEquation}. As time increases, the weight of the distribution shifts from the initial value close to 0 to the steady value at 3/4. (a) The initial operator is a single Pauli operator with size 1, and $q$ is 2. (b) Same as (a) but for a double Pauli operator with size 2. (c) The initial operator is a single Pauli operator with size 1, and $q$ is 3. The first peak in size distribution at small sizes is still present in the late time regime. (d) Same as (c) but for a double Pauli initial operator. }
    \label{fig:numericalSpin}
\end{figure}

\begin{figure}
    \centering
    \includegraphics{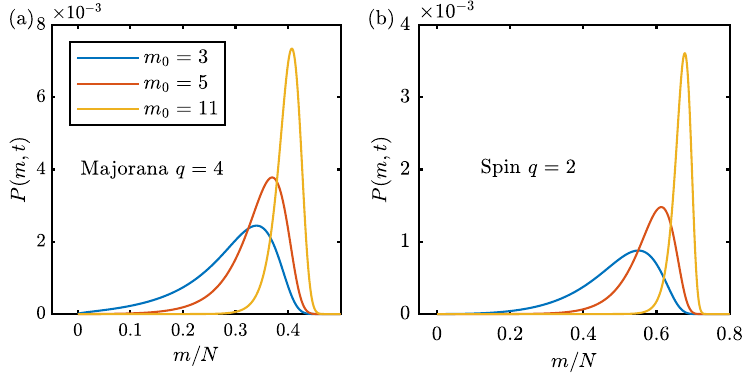}
    \caption{Size distribution at a fixed time for different initial sizes $m_0$. The system size $N=5000$. As the initial size increases, the distribution becomes narrower and centered around the mean value.}
    \label{fig:distrWidthNum}
\end{figure}

One can directly obtain the time-dependent operator size distribution by solving the master equations for very large system sizes
as the dimension scales linearly with $N$. The numerical results for $N=5000$ are presented in Fig.~\ref{fig:numericalMajorana} and Fig.~\ref{fig:numericalSpin} for the Majorana model and the spin model, respectively.~(In principle, one can solve the master equation for $N$ as large as $10^8$ by taking advantage of the sparse matrix and Krylov subspaces, or even larger using Monte Carlo sampling.) As time increases, the size distribution shifts away from the initial value and gradually converges to the steady distribution centered at $1/2$ for Majorana or $3/4$ for Pauli, where every Majorana~(Pauli) string becomes equally probable within their symmetry sector. During intermediate times, the operator size is characterized by a broad distribution. In the Majorana model, at $q=4$, the peak around the initial value survives for a long time for a single Majorana initial operator but decays quickly for a double Majorana initial operator. Conversely, at $q=6$, the initial peak persists for both initial conditions. In the spin model, the initial peak decays quickly for both initial conditions at $q=2$, but persists for a single Pauli initial operator at $q=3$. As the size of the initial operator increases, the width of the size distribution decreases for both Majorana and spin systems, shown in Fig.~\ref{fig:distrWidthNum}.

The numerical results indicate that the size distribution has a rich dependence on the initial operator and the value of $q$, with distinct differences between the Majorana and the spin model. These findings underscore the need for an analytical understanding of the size distribution for both Majoranas and spins. While a large $N$ solution has been derived for a single Majorana initial operator~\cite{zhang2023operator} and a single Pauli initial operator at $q=2$~\cite{zhou2019operator}, analytical solutions for general cases are still lacking.

\section{Solving the master equations in the large $N$ limit}
To explain the interesting features presented in the numerical results and to develop a systematic understanding of the operator size dynamics for general \(q\) and initial operators, we aim to develop an analytical solution to the master equation. Unfortunately, an exact solution is not feasible. Instead, we seek a solution valid in the large \(N\) limit.

Before going into the details, let us first outline our approach. Initially, the dynamics are governed by the large $N$ limit of the master equation with $m$ fixed. We will see that in this limit, the master equation simplifies significantly and can be solved exactly.
As time progresses, the probability distribution shifts toward larger sizes that scale with $N$. This shift occurs because the probability distribution is expected to peak at $N/2$ and $3N/4$ for the Majorana and the spin model, respectively. Therefore, beyond the early time regime, the appropriate large $N$ limit of the master equation involves fixing $m/N$ as $N$ approaches infinity. It is well-known that taking this large $N$ limit converts the master equation into a continuous Fokker-Planck equation, which can also be solved analytically in certain limits.
Thus, solving the master equation for initial operators with a fixed size involves two steps. First, we evolve the initial operator size distribution using the discrete master equation in the large $N$ limit. Then, we take the continuum limit of the evolved size distribution and continue evolving it using the Fokker-Planck equation, which governs the intermediate to late-time dynamics.

\subsection{Early time dynamics}
We first implement the first step and study the early time dynamics of the operator size distribution. One can take the large $N$ limit with fixed $m$ of the Majorana master equation in Eq.~\eqref{eq:MajoranaEquation} straightforwardly. In this limit, the transition rate $\gamma_\chi$ becomes,
\begin{equation}
   \lim\limits_{N\rightarrow\infty} \gamma^n_\chi(m) = \frac{1}{N^{n-1}}\frac{(q-1)!}{(q-n)!}\binom{m}{n}.
\end{equation}
Therefore, we only need to consider $\gamma_{\chi}^{(n=1)}(m)$, which corresponds to the transition rate to the largest possible size, $m+q-2$, from an operator of size $m$. Here, $\gamma_{\chi}^{(n=1)}(m) = m$. Transition rates to operators of smaller sizes are suppressed by large $N$.

For the spin model, in this large $N$ limit, only the term $n_z=n-1$ in Eq.~\eqref{eq:spinRate} survives, and the spin transition becomes
\begin{equation}
   \lim \limits_{N\rightarrow \infty} \gamma_\sigma^n(m) =  \frac{n}{(3N)^{n-1}}\frac{(q-1)!}{(q-n)!}\binom{m}{n}.
\end{equation}
It is proportional to the Majorana transition rates, and they are the same when $n=1$. Similar to the Majorana case, we only need to keep $\chi_\sigma^{(n=1)}$, the transition rate corresponding to the largest size increase $q-1$. 

Since the leading transition rates in the Majorana and the spin model are both $m$, the master equations are drastically simplified in this limit
\begin{equation}
\begin{aligned}
  &\partial_t P_\chi(m) = (m+2-q)P_\chi(m+2-q) - mP_\chi(m),\\
  &\partial_t P_s(m) = (m+1-q)P_s(m+1-q) -m P_s(m).
\end{aligned}
\end{equation}
This establishes the equivalence between the $q$-local Majorana model and the $(q-1)$-local spin model in the early time regime. One can unify the two equations as
\begin{equation}
\label{eq:earlyTimeMaster}
    \partial_t P(m) = (m-\lambda)P(m-\lambda) - m P(m),
\end{equation}
where $\lambda$ equals $q-2$ and $q-1$ for the Majorana and the spin model, respectively. This master equation was also studied in~\cite{yao2024notessolvable} for the Majorana model at $q=4$ and~\cite{zhou2019operator} for the spin model at $q=2$. As detailed in the appendix~\ref{sec:linearMaster}, one can solve this equation exactly by diagonalizing the matrix. The result takes the form of negative Binominal distribution,
\begin{equation}
    P(m,t)= \frac{\Gamma(m/\lambda)e^{-m_0 t}}{\Gamma(m_0/\lambda)\Gamma(\Delta M /\lambda +1)}(1-e^{-\lambda t})^{\Delta M/\lambda}\delta(\text{mod}(\Delta M,\lambda),0).
\end{equation}
where $m_0$ is the size of the initial operator, and $\Delta M = m-m_0$. Because of the connectivity of the master equation, $P(m,t)$ is nonzero only when $\Delta M$ is a multiple of $\lambda$. Since there is no backflow in Eq.~\eqref{eq:earlyTimeMaster}, $P(m,t)$ is zero when $m<m_0$ from the divergent denominator $\Gamma(\Delta M/\lambda +1)$.

After taking the continuum limit $m \rightarrow N x $ and then sending $N$ to infinity, the probability distribution function becomes,
\begin{equation}
    p(x,t) = \frac{N}{\lambda}\frac{e^{-m_0 t}}{(1-e^{-\lambda t})^{m_0/\lambda }\Gamma(m_0/\lambda)} (1-e^{-\lambda t})^{\frac{N x}{\lambda}}\left(\frac{Nx}{\lambda}\right)^{m_0/\lambda -1}.
\end{equation}
Recall that $\lambda$ equals $q-2$ for the Majorana model and $q-1$ for the spin model. When $m_0/\lambda <1$, the probability distribution function diverges at small $x$. This result indicates divergence in the size distribution for any initial operator with its size smaller than $q-2$ in the Majorana model or $(q-1)$ in the spin model. This fully agrees with the numerical results in Fig.~\ref{fig:numericalMajorana} and~\ref{fig:numericalSpin}. As we will see, this divergence persists beyond the early time regime. 

To proceed,  we define the scrambling time as
\begin{equation}
    t^* = \frac{1}{\lambda}\left(\log N -\log \lambda \right).
\end{equation}
Observe that at $t^*$, the probability distribution function in the large $N$ limit reduces to the $\chi$-squared distribution,
\begin{equation}
    p(x,t^*)=f_{m_0/\lambda}(x)=\frac{1}{\Gamma(m_0/\lambda)}e^{-x} x^{m_0/\lambda-1} 
\end{equation}
and is independent of $N$. However, recall that Eq.~\eqref{seq:earlyTimeMaster} is only valid when $m\ll N$, i.e., $x\ll 1$, while $p(x,t^*)$ is nonzero for the entire region of $x$ and is thus incorrect. To fix this, we need to consider the probability distribution function well before $t^*$,
\begin{equation}
\label{eq:earlyDistribution}
    p(x,t^*-\tau) = \frac{e^{\lambda \tau}}{\Gamma(m_0/\lambda)}e^{-e^{\lambda \tau }x} (e^{\lambda \tau }x)^{m_0/\lambda-1}  =e^{\lambda \tau }f_{m_0/\lambda}(e^{\lambda \tau }x)
\end{equation}
where $x$ is rescaled by a factor $e^{\lambda \tau}$. As long as $\tau \gg 0$, $p(x, t^*-\tau)$ only has weight on small $x$ and obeys the assumption for the early-time master equation.  Importantly, in the large $N$ limit with $\tau$ fixed, $p(x,t^*-\tau)$ becomes $N$ independent and can be further evolved by the Fokker-Planck equation to obtain the operator size distribution beyond the early time regime.

\subsection{Beyond early time}

As time increases, the operator with larger size density becomes more likely, and the approximation using Eq.~\eqref{eq:earlyTimeMaster} becomes invalid, and all the transitions must be considered. In this case, taking the continuum limit of the full master equation yields the Fokker-Planck equation,
\begin{equation}
    \partial_t p(x,t) =-\partial_x (v(x) p(x,t) )+ \frac{1}{N}\partial_x^2 (d(x) p(x,t))
\end{equation}
The first and the second terms are called drift and diffusion. In the large $N$ limit, we ignore the diffusion term and the Fokker-Planck equation can be solved analytically. We continue to evolve the early-time distribution in Eq.~\eqref{eq:earlyDistribution} using the Fokker-Planck equation and obtain the distribution at later times as
\begin{equation}
\label{eq:fokkerplanckEvolution}
    p(x,t)=\int dx_0 \delta(x-\bar{x}(t-t^*+\tau+\bar{x}^{-1}(x_0)))p(x_0,t^*-\tau)
\end{equation}
where $\bar x(t)$ obeys the first-order drift equation,
\begin{equation}
    \frac{d \bar x}{dt} = v(\bar x),
\end{equation}
and $\bar x^{-1}$ is the inverse function of $\bar x$.
Due to the $\delta$ function in the integral in Eq.~\eqref{eq:fokkerplanckEvolution}, the result is related to early-time probability distribution function $p(x, t^*-\tau)$ and thus the $\chi$-squared distribution by a change of variable,
\begin{equation}
\label{eq:generalP}
    p(x,t) = \frac{d z}{dx} f_{m_0/\lambda}(z) = \frac{dz}{dx}\frac{1}{\Gamma(m_0/\lambda)}e^{-z} z^{m_0/\lambda-1}.
\end{equation}
In other words, the distribution function at later times remains a $\chi$-squared distribution but of a new time-dependent variable $z$, which is related to $x$ as
\begin{equation}
\label{eq:generalZ}
    z = e^{\lambda \tau}\bar x(\bar x^{-1}(x)+t^*-\tau-t).
\end{equation}
We see that $z$ equals $e^{\lambda \tau} x$ when $t=t^*-\tau$, and the distribution reduces to the early-time distribution. 

\subsection{Full-time operator size distribution}
\begin{figure}
    \centering
    \includegraphics{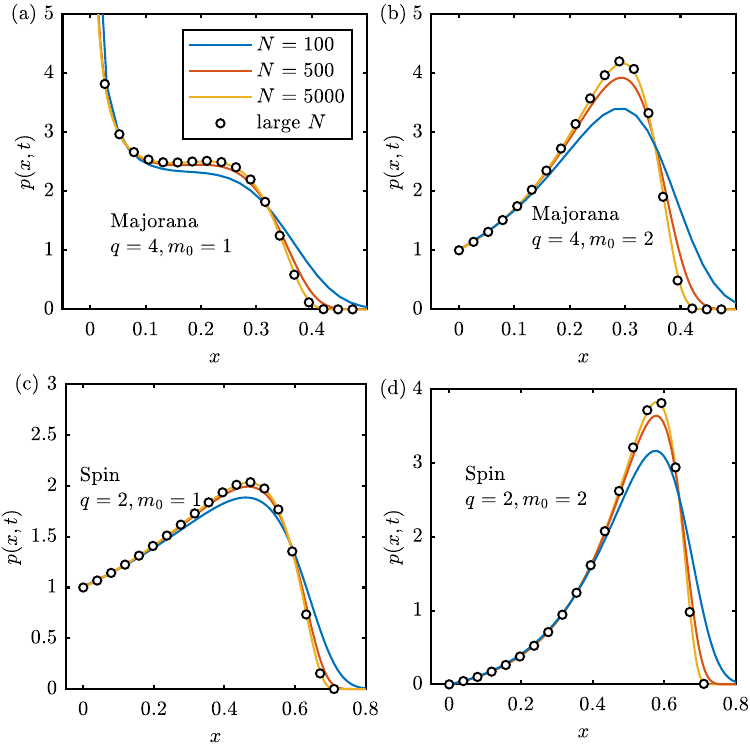}
    \caption{Comparing the large $N$ analytical solution with the numerical result from solving the master equations. In all cases, time is set to the scrambling time $t^*$. As $N$ increases, the numerical results become indistinguishable from the large $N$ solution. To match the discrete probability distribution $P(m)$ from numerics to the large $N$ continuous distribution function $p(x)$, $P(m)$ is multiplied by $N/2$ in the Majorana model at $q=4$ and by $N$ in the spin model at $q=2$.}
    \label{fig:theorynumerics}
\end{figure}
To apply the general result in Eq.~\eqref{eq:generalP} and Eq.~\eqref{eq:generalZ} to the Majorana and the spin model, we calculate the drift term in both models,
\begin{equation}
\begin{aligned}
   & v_\chi(x)  = \lim_{N\rightarrow\infty}\frac{1}{N}\sum_{n\in odd, 1\leq n<q}\gamma_\chi^{n}(N x)(q-2n), \\
    &v_\sigma(x)  = \lim_{N\rightarrow\infty}\frac{1}{N}\sum_{ 1\leq n\leq q}\gamma_\sigma^{n}(N x)(q+1-2n).
    \end{aligned}
\end{equation}
Remarkably, the drift term takes a unified simple expression for both models, as follows:
\begin{equation}
    v(x) = x^{eq}-x - x^{eq}\left(1-\frac{x}{x^{eq}}\right)^{\lambda+1}.
\end{equation}
Here $x^{eq}$ stands for the equilibrium of the mean size, which is $1/2$ for the Majorana model and $3/4$ for the spin model.

Solving drift equation $d \bar x /dt = v(\bar x)$, we get
\begin{equation}
    \bar x(t) = x^{eq}\left(1-\frac{1}{(1+e^{\lambda t})^{1/\lambda}}\right).
\end{equation}
In this case, in the large $\tau$ limit with $t-t^*$ fixed, the new variable $z$ is
\begin{equation}
    z = x^{eq}\frac{e^{\lambda(t^* - t)}}{\lambda}(-1+(1-x/x^{eq})^{-\lambda}).
\end{equation}
Plugging it into Eq.~\eqref{eq:generalP} gives rise to the unified operator size distribution in the large $N$ limit for $q$-local Majorana model and spin model.

As shown in Fig.~\ref{fig:theorynumerics}, the analytical result perfectly matches the numerical at large $N$, precisely characterizing the broad distributions for different initial operators. As the initial size $m_0$ increases, the $\chi$-squared distribution approaches a normal distribution peaked at the mean value $m_0/\lambda$ with width $\sim \sqrt{m_0/\lambda}$. Consequently, the size distribution $p(x,t)$ peaks at its mean value with width $m_0^{-1/2}$, matching the numerical results shown in Fig.~\ref{fig:distrWidthNum}.

Importantly, $z\approx e^{\lambda(t^*- t)} x$ at small $x$. From the $\chi$-squared distribution, $p(x,t)$ always diverges at small $x$ as $x^{m_0/\lambda-1}$ when $m_0<\lambda$. This divergence persists even at late time regimes.  In the Majorana model, the minimal value of $\lambda$ is 2, corresponding $q=4$. As a result, the Heisenberg time evolution of a single Majorana operator leads to at least square root divergence in the operator size distribution. At large $q$, the divergence becomes almost $1/x$. 

\begin{figure}
    \centering
    \includegraphics{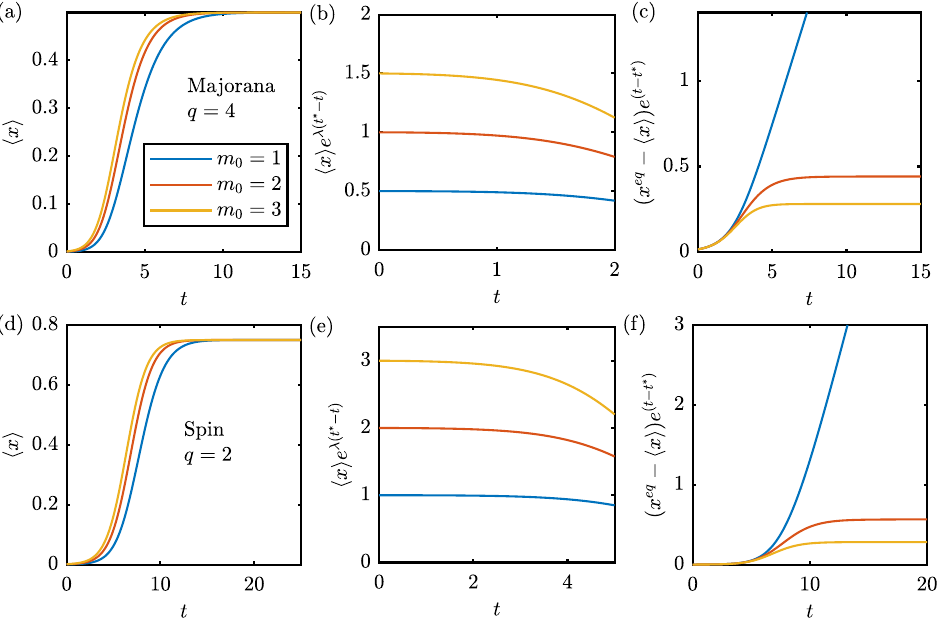}
    \caption{The mean size of Majorana operators as a function of time for various initial operator sizes $m_0$ and $q$. The results are obtained by solving the exact master equation numerically at $N=2000$.  In the early time, the mean size of all operators grows as $e^{\lambda t}$. In the late time, the mean size of a single Majorana/Pauli operator decays as $te^{-t}$ while all other operators decay as $e^{-t}$.}
    \label{fig:mean}
\end{figure}

We now consider the first moment of the size distribution, which is the mean size. It can be measured by the out-of-time ordered correlator as discussed in Sec.~\ref{sec:osd}. The mean size is
\begin{equation}
\label{eq:MajoranaMeanSize}
\langle x_{m_0}(t) \rangle = \int_0^{x^{eq}} p(x,t) x dx = \int_0^\infty f_{m_0/\lambda}(z)x(z,t) dz
\end{equation}
where 
\begin{equation}
    x = x^{eq}(1 -(1+\lambda e^{\lambda (t-t^*)}z/x^{eq})^{-1/\lambda}).
\end{equation}
When $t$ is well before the scrambling time $t^*$, $x\sim e^{\lambda(t-t^*)}z$.   In this regime, the integral in Eq.~\eqref{eq:MajoranaMeanSize} becomes
\begin{equation}
    x(t) = \frac{m_0}{\lambda}e^{\lambda (t-t^*)},
\end{equation}
showing the early-time exponential growth and $\lambda$ is the Lyapunov exponent.  This agrees with the numerical result from solving the master equation, shown in Fig.~\ref{fig:mean}~(b) and (e), where $\langle x \rangle$ by $e^{\lambda(t-t^*)}$ remains a constant $m_0/\lambda$ at small $t$.

On the other hand, when $t$ is much larger than $t^*$, $x\sim x^{eq} - x^{eq} e^{-(t-t^*)}(\lambda x/x^{eq})^{-1/\lambda}$.  As such, the integral exponentially decays to the steady value as $e^{-t}$, independent of $\lambda$. 
However, the case of $m_0=1$, corresponding to the single Majorana/Pauli initial operator, requires special treatment. In this case, $f_{m_0/\lambda}(z)x(z,t)$ diverges at small $z$ as $1/z$ and the integral does not converge.  This divergence is due to the naive large $t$ expansion of $x(z,t)$. Formally, one needs to keep all the orders of expansion and obtain a finite result by resummation. Instead, we calculate the integral exactly and get
\begin{equation}
\langle x_{1/\lambda}(t)\rangle = x^{eq}-x^{eq}(x^{eq}/\lambda)^{1/\lambda } U\left(\frac{1}{\lambda
   },1,\frac{e^{-\lambda(t-t^*) }x^{eq} }{ \lambda }\right)e^{-(t-t^*)}
\end{equation}
where $U$ is the confluent hypergeometric function~\cite{stanford2022subleading}.  Expanding the result at large $t$, we get
\begin{equation}
    \lim_{t\rightarrow\infty} \langle x_{1/\lambda}(t)\rangle=x^{eq}\left(1-\frac{\lambda}{\Gamma(1/\lambda)}\left(\frac{x^{eq}}{\lambda}\right)^{1/\lambda}te^{-(t-t^*)}\right)
\end{equation}
A single Majorana or a Pauli initial operator's mean size in the late time decays exponentially with a linear $t$ factor, while all other operators decay exponentially. This slow decay is confirmed by numerical results from simulating the exact master equation for $N=2000$. As shown in Fig.~\ref{fig:mean} (c) and (f),  $(x^{eq}-\langle x(t) \rangle)e^{t-t^*}$ approaches a straight line for a single Majorana/Pauli operator but a constant for other cases. We have also confirmed the late-time decay for other $q$.

For completeness, we also present the general result for $n$th moment of the distribution for arbitrary initial size $m_0$ and $\lambda$,
\begin{equation}
    \langle x_{m_0}^{n}(t) \rangle =\left(x^{eq}\right)^n \sum_k (-1)^k \binom{n}{k}a^{m_0/\lambda}U\left(\frac{m_0}{\lambda}, \frac{m_0-k}{\lambda}+1,a\right)
\end{equation}
where $a=e^{-\lambda(t-t^*)} x^{eq} /\lambda$. This generalizes the result of the Majorana model at $m_0=1$ in~\cite{stanford2022subleading, zhang2023operator} and the spin model at $q=2$, $m_0=1$ and $n=1$ in~\cite{zhou2019operator}. 

\section{Finite $N$ effects}
\label{sec:finiteN}
After understanding the full operator size distribution in the large $N$ limit, we briefly discuss some interesting finite 
$N$ effects in this section. These effects include deviations from the expected behavior observed in the large $N$ limit, such as fluctuations in the size distribution in the late time regime and the impact of the $q$ body interaction on the convergence of the numerical result to the large $N$ solution.

\subsection{Modulo-dependent branching within the exact symmetry sector}
\begin{figure}
    \centering
    \includegraphics{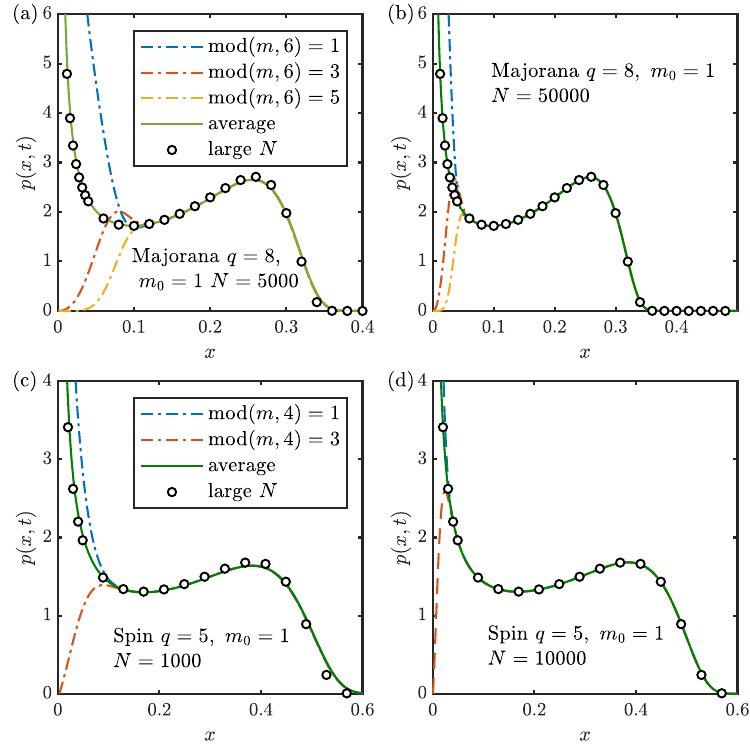}
    \caption{Modulo-dependent branching of the size distribution based on $\text{mod}(m,\lambda)$. The distribution within each branch converges to the large $N$ result much slower than their average. (a) and (b): The Majorana model at $q=8$. The master equation has two exact parity sectors, within which the size distribution further splits into three branches. (c) and (d): The Spin model at $q=5$. The master equation has two symmetry sectors from time reversal. Within each sector, the size distribution further splits into modulo-dependent branches. }
    \label{fig:sector}
\end{figure}
As discussed in Sec.~\ref{sec:master}, symmetries divide the master equation into multiple sectors for both Majoranas and spins. The Majorana master equation has two symmetry sectors for even $q/2$ and four for odd $q/2$. The spin master equation has one sector for even $q$ and two for odd $q$. These different sectors remain uncoupled at all times. Therefore one only needs to consider the size distribution within a particular sector determined by the initial size.

However, as shown in Fig.~\ref{fig:sector}, the size distribution, originating from the same initial size, further splits into multiple branches based on the modulo $\lambda$ relationship between $m$ and $m_0$, even within the same symmetry sector. For example, in the Majorana model at $q=8$~(Fig.~\ref{fig:sector}(a) and (b)), we consider the symmetry sector of the single majorana operator, corresponding to odd fermion parity and even under time reveal. The size distribution displays three distinct branches at small sizes that deviate from the large $N$ solution, which merge to the large $N$ solution at large sizes. As $N$ increases, the three branches slowly converge to the large $N$ result at small sizes. The spin model~(Fig.~\ref{fig:sector}(c) and (d)) also displays the same multi-branch structure.

The branching of the size distribution arises from the $q$-body interaction. At early times, the operator size distribution is governed by the master equation with linear transition rates defined in Eq.~\eqref{eq:earlyTimeMaster}, where size $m$ only transitions to size $m+\lambda$. Consequently, the early-time master equation splits into $\lambda$ independent sectors based on $\text{mod}(m, \lambda)$. For an initial operator with size $m_0$, the early-time discrete distribution $P(m,t)$ is nonzero if $m-m_0$ is a multiple of $\lambda$. As time progresses, the size distribution shifts away from the initial small sizes, these sectors start to couple unless forbidden by symmetries. The difference between these approximate sectors occurs at small sizes, leading to the branching behavior observed. The leading transition rate between these sectors scales with $N x^3$ for fermions and $N x^2$ for spins. Therefore, the branches are distinct at small $x$ but converge at large $x$.
Within each exact symmetry sector, the number of branches is $\lambda$ divided by the number of symmetry sectors.  

To capture the different branches in our analytical approach, one needs to derive the Fokker-Planck equation for each branch and couple these equations by the transition rates linear in $N$. However, the coupled equations are hard to solve. Instead, by adding these equations, the transition rates between the branches cancel out, resulting in the original Fokker-Planck equation, to which the large 
$N$ solution we derived applies. This corresponds to considering the average distribution for these branches. Indeed as shown in Fig~\ref{fig:sector}, the average distribution matches the large $N$ solution before the full distribution converges to it. The minimal $q$ for the branching is 8 for the Majorana model and 4 for the spin model.

\subsection{Late-time peak width}
\begin{figure}
    \centering
    \includegraphics{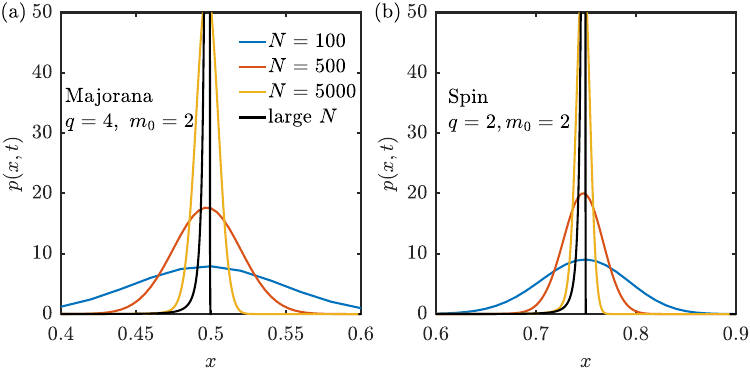}
    \caption{In the late time, the large $N$ size distribution approaches $\delta(x-x_{eq})$. The finite $N$ distribution, obtained from numerically solving the master equation, is a Gaussian distribution whose width decreases as $N$ increases. }
    \label{fig:peakwidth}
\end{figure}
In deriving the full distribution, we ignored the diffusion term in the Fokker-Planck equation. With only the drift term, the full size distribution in the long-time limit becomes a \(\delta\) function peaked at $x^{eq}$, while the actual steady distribution should be a binomial distribution at finite $N$. Therefore, in the late time, the finite $N$ numerical result converges to the large $N$ solution slowly, as shown in Fig.~\ref{fig:peakwidth}. To account for this finite $N$ effect, one needs to include the diffusion term in the Fokker-Planck equation, which broadens the $\delta$ distribution to a Gaussian distribution, thereby recovering the finite $N$ distribution. However, with the diffusion included, the Fokker-Planck equation no longer admits an analytical solution. Instead, one can adopt a variational approach and assume that the distribution for an initial $\delta$ function takes the form of a Gaussian distribution with a time-dependent mean and width, and solve the width as a function of the initial condition and time. The width should scale as $1/\sqrt{N}$.  We leave it to the future work. 

We also note that in the currently studied all-to-all connected models, the diffusion term, while broadening the size distribution, does not significantly affect the mean size. However, in higher dimensions, the diffusion term has a more dramatic effect and changes the shape of the wavefront~\cite{nahum2018operator,xu2019locality,khemani2018velocity,stanford2023scramblon}.

\section{Conclusion}
In this work, we present the exact master equations, numerical results, and the large $N$ size distribution characterizing the operator dynamics for the $q$-local Brownian Majorana SYK model and the spin model. Our results encompass the dynamics from the early-time regime to the late-time regime and are valid for arbitrary $q$ and initial operators. Remarkably, we show that the size distribution in both the Majorana model and the spin model is governed by the same distribution, which depends only on the early-time Lyapunov exponent $\lambda$, the late-time equilibrium size $x^{eq}$, and the initial size $m_0$. This distribution is related to the $\chi$-squared distribution through a change of variables. Through both numerical and analytical results, we reveal the universal features of the operator size distribution, including a broad distribution at intermediate times, divergence at small sizes for operators with initial sizes below a threshold related to $q$, and late-time slow decay for single Majorana/Pauli initial operators. We also discuss two major finite-$N$ effects, the modulo-dependent branching and late-time peak width,   and outline how to incorporate them into our analytical approach.
Although our study focuses on the Brownian model, we expect our results to apply to general all-to-all connected $q$-local quantum many-body systems at high temperatures. The broad size distribution and the strong initial size dependence have direct implications for high-temperature quantum many-body teleportation protocols~\cite{hayden2007black,schuster2021many}, which rely on a sharp distribution.

We also mention a few future directions. The derivation of the exact master equation relies on (anti)-commutation relations between different Majorana strings and Pauli strings. It would be interesting to construct other Brownian or random circuit models that can be described by a stochastic process. For example, if one imposes additional symmetry on the Brownian model, then the current approach does not apply anymore. Additionally, it would be valuable to develop a finite 
$N$ analytical solution for the size distribution following the approach discussed in Sec.~\ref{sec:finiteN}. 
Furthermore, the finite temperature effects on the size distribution~\cite{qi2019quantum, lucas2019operator} are important, especially the fate of the broad distribution and the divergence at small sizes. By combining the early-time discrete master equation and intermediate-time continuous Fokker-Planck equation, we show that the full-time distribution is related to the early-time distribution by a change of variables, and the new variable only depends on the drift term in the Fokker-Planck equation. Using a completely different approach, dubbed the "two-way" approach~\cite{gu2022twoway}, recent work~\cite{zhang2023operator} shows that the size distribution in the large $N$ static SYK model with energy conservation also has the same structure, where the early-time distribution in their approach is related to the retarded vertex function and the new variable $z$ is related to the advanced vertex function. By comparing the two approaches, one may derive an approximate master equation to describe the operator dynamics in static systems and study how finite temperature enters the size distribution. More generally, it would be interesting to think about the definition of operator size in systems with additional symmetry~\cite{chen2020many} and compare it with $K$-complexity~\cite{parker2019universal}.

\section{Acknowledgements} S. Xu thanks helpful discussion with Yingfei Gu, Shunyu Yao, Lakshya Agarwal, Shao-Kai Jian, Michael Winer, and Brian Swingle. S. Xu thanks Shunyu Yao for coordinating the submission of the related work~\cite{yao2024notessolvable}, which will appear in the same arXiv posting. S. Xu acknowledges the support from the Google Research Scholar Program and the advanced computing resources provided by Texas A\&M High Performance Research Computing.

\appendix
\begin{appendices}
\section{Solving the master equation in the early time regime for arbitrary initial condition}
\label{sec:linearMaster}
For an initial operator of a fixed size, the large $N$ limit of the master equations for both the Majorana model in Eq.~\eqref{eq:MajoranaEquation} and the spin model in Eq.~\eqref{eq:spinEquation} take the following form,
\begin{equation}
\label{seq:earlyTimeMaster}
    \partial_t P(m) = (m-\lambda)P(m-\lambda) -  m P(m),
\end{equation}
where $\lambda$ equals $q-2$ and $q-1$ for the Majorana model and the spin model respectively. 
To solve the equation, we rewrite it in the matrix form,
\begin{equation}
    \partial_t \vec P = A\vec P, A_{m,m'}=m'(\delta_{m+\lambda,m'}-\delta_{m,m'}).
\end{equation}
The matrix is lower triangular and its eigenvalues are its diagonal entries, which are non-negative integers $k$. The corresponding left and right eigenvectors are,
\begin{equation}
\begin{aligned}
&v_k(m) = \frac{(-1)^{(m-k)/\lambda}\Gamma(k/\lambda)}{\Gamma(1+(k-m)/\lambda)\Gamma(m/\lambda)}\delta(\text{mod}(m-k,\lambda),0),\\
&u_k(m) = \frac{\Gamma(m/\lambda)}{\Gamma(1+(m-k)/\lambda)\Gamma(k/\lambda)}\delta(\text{mod}(m-k,\lambda),0).
\end{aligned}
\end{equation}
The left/right eigenvectors are nonzero when $(m-k)$ is a multiple of $\lambda$ since the master equation only connects the component $p(m)$ to $p(m+\lambda)$. Furthermore, the left eigenvector $v_k(m)$ vanishes when $m > k$  and the right eigenvector $u_k(m)$ vanishes when $m < k$. One can check left eigenvectors and the right eigenvectors with different eigenvalues are orthogonal,
\begin{equation}
    \sum_{m=0}^\infty v_{k'}(m) u_{k}(m) =\frac{\Gamma(k'/\lambda)\delta(\text{mod}(k'-k,\lambda),0)}{\Gamma(k/\lambda)\Gamma((k'-k)/\lambda+1)}\sum_{l=0}^{(k'-k)/\lambda}(-1)^l\binom{(k'-k)/\lambda}{l}=\delta_{k',k}. 
\end{equation}
The solution of the equation can be written as
\begin{equation}
    P(m,t) = \sum_{m',k} P(m',0)v_k(m') u_k(m)e^{-k t} 
\end{equation}
Consider the initial distribution has a fixed size $m_0$. One can sum the series analytically and get
\begin{equation}
\begin{aligned}
   P(m,t)& = \sum_{k\geq m_0} v_k(m_0)u_k(m)e^{-kt}\\ &= \frac{\Gamma(m/\lambda)e^{-m_0 t}}{\Gamma(m_0/\lambda)\Gamma((m-m_0)/\lambda +1)}(1-e^{-\lambda t})^{(m-m_0)/\lambda}\delta(\text{mod}(m-m_0,\lambda),0).
\end{aligned}
\end{equation}
Because of the connectivity of the master equation, $P(m,t)$ is nonzero only when $m-m_0$ is a multiple of $\lambda$. Since there is no backflow in Eq.~\eqref{seq:earlyTimeMaster}, $P(m,t)$ is zero when $m<m_0$ from the divergent denominator $\Gamma((m-m_0)/\lambda +1)$.
\end{appendices}
\bibliography{main}

\end{document}